\documentclass[11pt,a4paper]{article}
\pdfoutput=1
\usepackage[english]{babel}
\usepackage{jheppub}

\newcommand{\beq}{\begin{equation}}
\newcommand{\eeq}{\end{equation}}
\newcommand{\bea}{\begin{eqnarray}}
\newcommand{\eea}{\end{eqnarray}}

\title{Charged black rings in supergravity with a single non-zero gauge field}
\author{Andrew Feldman}
\author{and Andrei A. Pomeransky}
\affiliation{Budker Institute of Nuclear Physics,\\
11, academician Lavrentiev Ave., Novosibirsk, Russia}
\affiliation{Physics Department, Novosibirsk State University,\\
2 Pirogova Str., Novosibirsk, Russia}
\emailAdd{andrew.l.feldman@gmail.com}
\emailAdd{pomeransky@gmail.com}
\abstract{General charged black ring solution with two angular momenta, a charge and a dipole charge is found by the inverse scattering method. The solution is presented in a relatively concise form in which its symmetries are manifest. The regularity conditions are found and the physical characteristics of the regular solution are expressed via its parameters.}
\arxivnumber{1206.1026}

\begin{document}

\maketitle

\flushbottom

\selectlanguage{english}

\section{Introduction}
Black holes are one of the most interesting types of exact
solutions in general relativity. In 4D space-time the black hole solutions are by now investigated rather completely. The completeness of our knowledge in this case is  due to the uniqueness theorem, stating that there is only Kerr-Newman solution and some its direct generalizations. They all have an event horizon with spherical topology. The situation is very different in the higher dimensional case. Here one has a bigger variety of black hole types. The first known example of a black hole with an event horizon of non-spherical topology was the  Emparan-Reall solution \cite{Emparan:2001}. It was called black ring due to
having the $S^1\times S^2$ event horizon topology. The black rings were soon generalized to the case of charged solutions in supergravity (see \cite{Emparan:2006,Emparan:2008} for a review). It was noted, that in general a black ring has two independent charges for each abelian gauge field in the theory. One of them is the usual charge that can be found from the Gauss theorem. The other charge can be calculated by integrating the field strength 2-form  over a closed surface encircling the event horizon. It is called dipole charge for historical reasons.

In spite of these efforts, the families of charged solutions found so far were not general enough. The general {\emph neutral} regular black ring was found in \cite{Pomeransky:2006}. It has two independent angular momenta parameters, corresponding to rotations in two mutually orthogonal planes. It was found by the inverse scattering method \cite{Belinsky:1978,Belinsky:1979}, adapted to
the higher-dimensional case in \cite{Pomeransky:2005}. Finding the general charged solution is a considerably more difficult problem. However, in the case of a single non-zero abelian gauge field it can be solved by the same method. The only difference is that now one has to consider pure gravity in 6D space-time \cite{Rocha:2011} (see also \cite{Yazadjiev:2006a,Yazadjiev:2006b}). It is well-known from the Kaluza-Klein theory, that a gauge field and a scalar dilaton arise here after compactification of the fifth spacial coordinate.

This approach has allowed to find  recently a very important
solution with a non-zero dipole charge, but still with a zero
usual charge \cite{Chen:2012}. It is worth to note that the
regular solution was presented in \cite{Chen:2012} in a relatively simple and compact form, which is always a difficult task. An analogous family of solutions was found in \cite{Rocha:2012}, which has both non-zero dipole and usual charges. However, as noted by  the authors of \cite{Rocha:2012}, the two charges  in their family of solutions are not independent. In this article we
present the general black ring solution with independent usual charge and dipole charge with respect to a single non-zero abelian gauge field. The general solution has unexpectedly many symmetries. These symmetries allow us to present it in a
relatively concise form.  In a certain sense, the general solution looks simpler and is  more tractable than its particular cases. The  vanishing of 6D Ricci tensor is demonstrated by showing that the tensor components are equal to divergence of a curl. Imposing the conditions of absence  of Dirac-Misner and conical singularities in the 5D metrics allows one to find the general regular solution. For this regular solution we calculate mass, two angular momenta and angular velocities, event horizon volume and temperature, the two charges and the corresponding potentials. Then we check that the First law of black hole mechanics holds for
these quantities.

\section{Derivation}
We use the Inverse scattering method (ISM) as described in \cite{Belinsky:1978,Belinsky:1979,Belinski:2001,Pomeransky:2005,Emparan:2008}.
Let us recall some basic equations of ISM in order to fix our notations. The metric in our case can be reduced to the block-diagonal form:
\begin{equation} \label{metric_blocks}
ds^2=g_{A B}(\rho, z)dx^A dx^B+f(\rho, z)(dz^2+d\rho^2), \quad A,B=t,\phi,\psi,w\,.
\end{equation}
\noindent
Let us denote $\hat{g}$ the matrix $g_{A B}$ with $A,B=t,\phi,\psi,w$.
The field equations coincide with vanishing of the 6D Ricci tensor. This is equivalent to
\begin{equation}
\label{Ricci}
\partial_{i} g^{A B} \sqrt{h} h^{i j} \sqrt{-g} \partial_{j} g_{B C}=0, \quad i,j=\rho,z\,.
\end{equation}
The conformal factor $f(\rho, z)$ satisfies the equations:
\begin{equation} \label{Conffactor}
\partial_{\rho}(lnf)=-\frac{1}{\rho}+\frac{1}{4 \rho} Tr(U^2 - V^2), \quad \partial_{z}(lnf) = \frac{1}{2 \rho} Tr(UV).
\end{equation}
ISM allows one starting from a solution $g_{0\,AB}$ to get a new solution $g_{AB}$. The explicit relation between them is
\begin{equation}
g_{ab}=g_{0\,ab}-\sum_{k,l} D^{kl}\nu_k^{-1}\nu_l^{-1}N^{(k)}_a N^{(l)}_b\,,
\end{equation}
where
\begin{eqnarray} \label{mnvectors}
m^{(k)a} = m^{(k)}_{0b} [\psi_0^{-1}(\nu_k,\rho,z)]^{ba}, \quad n^{(k)}_a = \sum_l \nu_l^{-1}D^{kl}N^{(l)}_a\,.
\end{eqnarray}
The matrix $\psi_0^{-1}$ is the solution of a system of linear
differential equations (for details, see the references on ISM
cited above). The functions $\nu_k$ have the form $\nu_k=w_k-z \pm
\sqrt{(w_k-z)^2+\rho^2}$, where the plus corresponds to a soliton, and the minus  to an anti-soliton, $w_k$ are constants. The notations $\Gamma_{kl}=m^{(k)a}g_{0\,ab}m^{(l)b}
(\rho^2+\nu_k\nu_l)^{-1}$, $N_a^{(l)} = m^{(l)c} g_{0\,ca},$ were also introduced here, with the vectors $m^{(k)}_{0a}$ been composed of arbitrary constants. $D^{kl}$  is the inverse matrix
for $\Gamma_{kl}$: $D^{km}\Gamma_{ml} = \delta^k_l$. The range of
the values of indices $k$, $l$ and $m$ is the number of solitons
added to the seed solution $g_{0 a b}$. The number of solitons
used in this work is 3.

It is convenient to use coordinates $(u,v)$, which makes all metric components rational functions. These coordinates are related to the original  $(\rho, z)$ as follows:
\begin{equation} 
\rho^2=-\frac{4k^4 (1-u^2)(1+cu)(1-v^2)(1+cv)}{(u-v)^4}, \quad z=\frac{k^2(1-u v)(2+cu+cv)}{(u-v)^2}.
\end{equation}
We use the following seed metrics:
\begin{equation}
\label{br1}
ds^2 = A(dt+\omega_t d\phi)^2-\frac{\mu_1\mu_3}{\mu_2 A}\,d\phi^2-B(dw+\omega_w d\psi)^2+\frac{\mu_1\mu_3}{B \mu_2}\,d\psi^2-f(d\rho^2+dz^2)\,,
\end{equation}
where
\begin{equation} A=(1+bv)/(1+bu), \quad B=(1-\mu v)/(1-\mu u), \end{equation}
\begin{equation} \omega_t=\sqrt{\frac{2b(b-c)(1+b)}{(1-b)}}\frac{k(1+v)}{1+b v}, \quad \omega_w=\sqrt{\frac{2\mu(\mu+c)(1-\mu)}{(1+\mu)}}\frac{k(1+v)}{1-\mu v}.
\end{equation}
The functions  $\mu_i$  have the following form in the coordinates $\rho$ and $z$:
\begin{equation} \mu_i=w_i-z+\sqrt{(w_i-z)^2+\rho^2}, \ \ \ i=1,2,3,
\end{equation}
\begin{equation} w_1=-c k^2, \ \ \ w_2=c k^2, \ \ \ w_3=k^2\,.
\end{equation}
\noindent
While in coordinates $(u,v)$ they are
\begin{equation} \mu_1=-\frac{2k^2(1-u)(1+v)(1+cv)}{(u-v)^2}, 
\end{equation}
\begin{equation}
\mu_2 = -\frac{2k^2(1-u)(1+v)(1+cu)}{(u-v)^2},
\end{equation} \begin{equation} \mu_3 = -\frac{2k^2(1-v^2)(1+cu)}{(u-v)^2}.
\end{equation}
Explicitly, the seed matrix $\hat{g}_0$ is
\begin{equation} \label{seed} \hat{g}_0=\left( \begin{array}{cccc} A & A\omega_t & 0 & 0 \\ A\omega_t & A\omega_t^2-\frac{\mu_1\mu_3}{\mu_2 A} & 0 & 0 \\ 0 & 0 & -B\omega_w^2+\frac{\mu_1\mu_3}{\mu_2 B} & -B\omega_w \\ 0 & 0 & -B\omega_w & -B \\ \end{array} \right)
\end{equation}
It is block-diagonal. Its upper-left block coincides with the
corresponding block in the Emparan-Reall metric. The other block is essentially the same. It can be obtained from the first one by changing the overall sign and exchanging $t\leftrightarrow w$, $\phi\leftrightarrow \psi$ and $b \leftrightarrow  -\mu$.  The seed matrix determinant is  $\det{\hat{g}_0}=\mu_1^2 \mu_3^2/\mu_2^2 \neq\rho^2$. The matrix $\psi_0^{-1}$ that corresponds  to this $\hat{g}_0$ also has the block-diagonal form and can be presented as the product:
$\psi_0^{-1}=\psi_{-1}^{-1}\chi_0^{-1}$, where
\begin{equation}
\chi_0^{-1}=\left( \begin{array}{cccc} \lambda - \frac{\rho^2}{A \mu_2} + \frac{A \rho^2 \omega_t^2}{\mu_1\mu_3} & -\frac{A \rho^2 \omega_t}{\mu_1\mu_3} & 0 & 0\\ -A \mu_2 \omega_t & \lambda + A \mu_2 & 0 & 0\\ 0 & 0 & \lambda + B \mu_2 & - B \mu_2 \omega_w\\ 0 & 0 & \frac{-B \rho^2 \omega_w}{\mu_1\mu_3} & \lambda - \frac{\rho^2}{B \mu_2} + \frac{B \rho^2 \omega^2}{\mu_1\mu_3}\\ \end{array} \right)\,,
\end{equation}
\beq \psi_{-1}^{-1}=diag\Biggl(-\frac{\mu_2 - \lambda}{\lambda}, -\frac{(\rho^2/\mu_1+\lambda) (\rho^2/\mu_3 + \lambda)}{\lambda^2}, \frac{(\rho^2/\mu_1+\lambda) (\rho^2/\mu_3 + \lambda)}{\lambda^2},\frac{\mu_2 - \lambda}{\lambda}\Biggr).
\eeq
Since the blocks of the seed metric appear also in the Emparan-Reall metric, the corresponding $\psi_0^{-1}$ can be extracted from the  ISM derivation of Emparan-Reall black ring \cite{Iguchi:2006,Tomizawa:2006}, and its $2\times 2$ block was already used in \cite{Pomeransky:2006}.

To this seed background we add two solitons and one anti-soliton: we take the functions $\nu_k$ equal to $\nu_1=\mu_1$, $\nu_2=-\rho^2/\mu_2$, $\nu_3=\mu_3$. The determinant of the resulting matrix is $\det \hat{g}=\rho^2$ as it should be. The constants $m^{(k)}_{0a}$ can be taken such that $m^{(1)}_{0t}=m^{(1)}_{0w}=m^{(2)}_{0\phi}=m^{(2)}_{0\psi}=m^{(3)}_{0t}=m^{(3)}_{0w}=0$. With these components non-zero the resulting solution would have more complicated form and additional singularities would appear.

\section{Metrics}
The metric derived in the previous section can be reduced by some transformations of coordinates and parameters to the following maximally symmetric and compact form:
\bea 
\label{metrica6D}
ds^2&=&-\frac{D(v,u)}{H(u,v)}\left(dt+\Omega^t
\right)^2+2\frac{K(u,v)}{H(u,v)}\left(dt+\Omega^t \right)
\left(dw+\Omega^w \right)\\\nonumber &+&\frac{D(u,v)}{H(u,v)}
\left(dw+\Omega^w \right)^2+\frac{F(u,v)}{G(y_1)
H(v,u)}d\phi^2+2\frac{J(u,v)}{H(v,u)} d\phi d\psi \\ \nonumber &-&
\frac{F(v,u)}{G(y_2) H(v,u)}d\psi^2 -\frac{2 k^2 H(u,v)}{c(u-v)^2}
\left(\frac{du^2}{G(u)}-\frac{dv^2}{G(v)} \right).
\eea
The metric depends only on two coordinates, $u$ and $v$. It is independent of time $t$, Kaluza-Klein coordinate $w$  and angles $\phi$ and $\psi$. The metric depends also on constant parameters $x_i$, $a_i$, $i=1,2,3$ and  $y_\alpha$, $\alpha=1,2$. We have chosen to denote the arbitrary constant in the conformal factor as $k^2/c$ for future convenience. Several functions enter the expression (\ref{metrica6D}). The function $H(u,v)$ appears in the denominator of all components of metric and inverse metric. It can be expressed as \beq\label{H}
H(u,v)=S\{-(u-y_1)(v-y_2)\} + 2 a_1 a_2 a_3 r_1 r_2 r_3 \sum_i
\frac{C_i}{a_i r_i}, \eeq where the combinations of constants were introduced:
\beq\label{Cs}
C_i= 1 - \frac{a_i^2 }{(y_1-x_i)(y_2-x_i)}\,.
\eeq
The functions $r_i$ are defined as
\beq r_i
=\frac{(u-x_i)(v-x_i)}{\frac{dG}{du}\vert_{u=x_i}}-1 \,, 
\eeq
where  $G(u)=(u-x_1)(u-x_2)(u-x_3)$ is a cubic polynomial, which has $x_i$ as its roots. The operator $S\{\dots\}$ is the
composition of operators
$S\{f(u,v)\}=S_1\{S_2\{S_3\{f(u,v)\}\}\}$, which act on functions of
coordinates  $u$ and $v$ in the following way: \beq
S_i\{f(u,v)\}=f(u,v) - \frac{a_i^2 }{(y_1-x_i)(y_2-x_i)}\
\frac{(u-v)^2}{(h_i(u)-h_i(v))^2} f(h_i(v),h_i(u)). \eeq Here
$h_i$ are M\"{o}bius transformations: \beq h_i(u)
=\frac{\frac{dG}{du}\vert_{u=x_i}}{u-x_i}+x_i\,. \eeq The
functions $D(u,v)$ and  $K(u,v)$ are given by the following
relations: \bea\label{D} D(u,v)&=& S\{-(u-y_1)(u-y_2)\} + 2r_1 r_2
r_3 a_1 a_2 a_3  \sum_i \frac{C_i}{a_i r_i}\,,\\\nonumber
D(u,v)&=&H(u,v)+ S\{(u-v)(y_1-u)\}\,,\;D(v,u)=H(u,v)+
S\{(u-v)(y_2-v)\}\,, \eea
\beq\label{K} K(u,v)= -(u-v)(a_1 r_1 C_2
C_3+ a_2 r_2 C_1 C_3+ a_3 r_3 C_1 C_2).
\eeq
Functions $F(u,v)$ and $J(u,v)$ have the form:
\beq\label{F}
F(u,v)=S\left\{-\frac{G(v)}{(u-v)^2}(u-y_1)(u-y_2)\right\} + 2 \frac{ a_1 a_2
a_3}{(u-v)^2}  \sum_i \frac{r_1 r_2 r_3}{a_i r_i}
S_i\left\{\frac{G(v)(u-x_i)}{(v-x_i)}\right\}\,, \eeq \beq\label{J} J(u,v)=
\sum_i \frac{a_i r_i}{(y_1-x_i)(y_2-x_i)(u-v)}
S^\prime_i\{(u-x_i)(v-x_i)\} -\frac{\eta a_1 a_2 a_3 r_1 r_2 r_3
}{G(y_1)G(y_2)(u-v)}\,,
\eeq
where a combination of constants is introduced:
$$\eta=(x_2-x_3)^2(x_1-y_1)(x_1-y_2)+(x_1-x_3)^2(x_2-y_1)(x_2-y_2)+(x_1-x_2)^2(x_3-y_1)(x_3-y_2).$$
Here and in what follows $S^\prime_i$ means the composition of all
$S_1$, $S_2$, $S_3$, except $S_i$. For example,
$S^\prime_1\{f\}=S_2\{S_3\{f\}\}$. Finally, the differential form
$\Omega^t$ has the form:
\bea\label{Omega}
\Omega^t &=& \frac{d\phi}{H(v,u)}\left(S\{y_2-u\}+ a_1 a_2 a_3 r_1 r_2 r_3 \sum_i
\frac{C_i}{a_i r_i}\left(\frac{1}{y_2-x_i}-\sum_j \frac{1}{y_2-x_j}\right)\right)\\
\nonumber &+& \frac{d\psi}{H(v,u)} \sum_i \frac{a_i
r_i}{(y_2-x_i)(u-v)} S^\prime_i\{(u-v)(u-x_i)\}\,.
\eea
The other differential form in (\ref{metrica6D}), $\Omega^w$, can be
obtained from  $\Omega^t$ by exchanging $u\leftrightarrow v$,
$y_1\leftrightarrow y_2$ and $d\phi \leftrightarrow d\psi$.

While (\ref{metrica6D}) is probably the simplest way to present the solution, the explicit list of the metric components is not much more complicated. We give it in the Appendix. The inverse metric tensor is  given there as well.
Metric  (\ref{metrica6D}) is Ricci flat and thus it satisfies Eq. (\ref{Ricci}).  To demonstrate this it is useful to note that  (\ref{Ricci}) is equivalent to the existence of a matrix ${\cal N}^A_C$, such that
\begin{equation}\label{Ricci_flat}
g^{A B} \sqrt{h} h^{i j} \sqrt{-g} \partial_{j} g_{B C}=\epsilon^{i j} \partial_{j} {\cal N}^A_C,
\end{equation}
where $\epsilon^{i j}$ is a constant anti-symmetric matrix, $\epsilon^{i j}=-\epsilon^{j i}$. We were able to find
${\cal N}^A_C$ explicitly for this metric. It has rather simple form and is presented in the Appendix. Apart from allowing to check the Ricci flatness, the matrix $\cal{N}^A_C$ is also very useful for calculating duals of Kaluza-Klein gauge fields arising after compactifications \cite{Gal'tsov:2008,Gal'tsov:2009}.

Finally, let us explain the origin of the operators $S_i\{\dots\}$. Their appearance is due to existence of three symmetries of the metric with respect to certain M\"{o}bius transformations of coordinates $u$ and $v$. Namely, the transformation: $u\rightarrow h_i(v)$, $v \rightarrow h_i(u)$, $a_i \rightarrow (y_1-x_i)(y_2-x_i)/a_i$ does not change metric components. This is true also for the inverse metric and ${\cal N}^A_C$. It is clear now that the operators $S_i\{\dots\}$ allows one to construct expressions that possess this invariance. One can speculate that analogous symmetries will remain intact in the solution with 3 non-zero gauge fields which is still to be found.

\section{Noether charges and other quantities}
After compactification on a circle the solution (\ref{metrica6D}) in general gives a 5D metric with conical and Dirac-Misner string singularities. To get rid of the singularities one has to impose some constraints. The convenient way to analyze the singularities is to consider the rod structure \cite{Harmark:2004,Emparan:2008} of the metric. The rods are segments of the $\rho=0$ axis, and the rod directions are the eigenvectors of metric at these segments with zero eigenvalue. These directions can be most efficiently calculated in practice from the residues of the corresponding poles in the inverse metric. We will not describe here the complete procedure of the eliminating of singularities as it is well-known. We will state only resulting regularity conditions for the parameters instead.

In this section we will use the following particular choice of parameters: $x_1=1$, $x_2=-1$, $x_3=-1/c$, $y_1=-1/b$, $y_2=1/\mu$. First of all, in order  to get the regular solution one should perform a boost in the $w$ direction with the speed equal to $\beta=\frac{b c}{b-c}a_3$. One has to impose also two other conditions $a_2=-a_1 \frac{1-c}{1+c}$ and
\begin{equation} \label{m b c regularity} \frac{1+b}{1-b} \frac{1-\mu}{1+\mu}=\left(\frac{1+c}{1-c}\right)^2\,.
\end{equation}

After the regular solution is found, one can calculate such quantities as mass, charges and corresponding potentials, angular momenta, horizon volume, temperature and angular velocities. To write down the results in a compact form we need to introduce notations for the following set of constants:
\begin{equation} \label{first delta}
\Delta_1=\left(1+\frac{b \mu}{(1+b) (1-\mu)} a_1^2 \right), \quad \Delta_2=\left(1-\frac{ b \mu}{(1+b) (1-\mu )} a_1^2\right), \end{equation}
\begin{equation} \Delta_3=\left(1-\frac{1-c}{1+c}\frac{b \mu}{(1+b)(1+\mu)} a_1^2\right), \quad \Delta_4=\left(1-\frac{(c+\mu ) b \mu }{(b-c)(1-\mu )^2} a_1^2\right),
\end{equation}
\begin{equation}
\Delta_5=\left(1-\frac{c^2 b \mu}{(b-c) (c+\mu)} a_3^2\right), \quad \Delta_6=\left(1-\frac{c^2 b^2}{(b-c)^2} a_3^2\right), \end{equation}
\begin{equation}
\Delta_7=\left(1-\frac{c b \mu}{(b-c)(1-\mu )} a_1 a_3\right), \quad \Delta_8=\left(1+\frac{1-c}{1+c}\frac{c b \mu}{(b-c)(1+\mu)} a_1 a_3\right),
\end{equation}
\begin{equation} \label{last delta}
\Delta_9=\left(1+\frac{1-c}{1+c}\frac{(c+\mu) b^2 c}{(b-c)^2(1+\mu)} a_1 a_3\right), \quad \Delta_{10}=\left(a_1-\frac{(1-\mu) c}{(c+\mu)} a_3\right),
\end{equation}
\noindent
where the parameters $\Delta_1$ and $\Delta_5$ coincide in the regular case with $C_1$ and $C_3$, which were introduced in ($\ref{Cs}$).

The mass can be read off from the asymptotics \cite{Harmark:2004}:
\begin{equation}
g_{tt}=-1+\frac{8 M}{3 \pi r^2}+{\cal O}\left(\frac{1}{r^4}\right).
\end{equation}
Using the notations $(\ref{first delta})-(\ref{last delta})$ one can write down the following expression for the mass:
\bea
M &=& \frac{\pi k^2 (1+c \mu) (1+\mu) (b-c)}{(1-c) \left(1-c^2\right) \left(c+\mu \right)}\frac{\Delta_1}{\Delta_6}\Biggl(\frac{\mu \left(1-2 c^2 \right)-c}{(1+c \mu )}\Delta_3 \Delta_5 \Delta_6 \\ \nonumber
&+& \frac{4 c (c+\mu)}{\left(1-c^2\right) \mu }\Delta_3 \Delta_5-\frac{4 c^2\text{ }(1+c \mu )}{ \left(1-c^2\right) \mu }\Delta_3+\frac{(1-c) \mu}{(1+\mu)} \Delta_2 \Delta_6^2\Biggr).
\eea
For the angular momentum $J_{\phi}$ one has at $r\rightarrow \infty$:
\begin{equation} g_{t \phi}=-\frac{4 J_{\phi}}{\pi} \frac{sin^2 {\theta}}{r^2} \left(1+{\cal O}\left(\frac{1}{r^2}\right)\right),
\end{equation}
\noindent{and for the other component $J_{\psi}$}:
\begin{equation} g_{t \psi}=-\frac{4 J_{\psi}}{\pi} \frac{cos^2 {\theta}}{r^2}\left(1+{\cal O}\left(\frac{1}{r^2}\right)\right),
\end{equation}
\noindent
where  the angle $\theta$ is defined so that at $r\rightarrow \infty$:
\begin{equation} g_{\phi \phi}=r^2 sin^2 {\theta}\left(1+{\cal O}\left(\frac{1}{r^2}\right)\right),
\end{equation}
\begin{equation}
g_{\psi \psi}=r^2 cos^2 {\theta}\left(1+{\cal O}\left(\frac{1}{r^2}\right)\right).
\end{equation}
The resulting expressions  for $J_{\phi}$ and $J_{\psi}$ has the form:
\begin{equation}
J_{\phi} = \frac{\sqrt{2} \pi k^3 (1+b) (1+\mu) \sqrt{b (c+\mu)}}{(1+c)^{3/2} \sqrt{1-c}} \frac{\Delta_1 \Delta_3 \Delta_5 \left[\Delta_3+\Delta_5-2 \Delta_8+\left(\frac{1+c}{1-c}\right) \Delta_8^2\right]}{\sqrt{\Delta_6}},
\end{equation}
\begin{equation}
J_{\psi}=\frac{2 \sqrt{2} \pi k^3 c b (1+\mu ) \sqrt{\mu (b-c)}}{\left(1-c^2\right)^{3/2}}\frac{\Delta_1 \Delta_3 \Delta_5 \Delta_7 \Delta_{10}}{\sqrt{\Delta_6}}.
\end{equation}
The electric charge $q$ can be defined in a coordinate-independent way as \cite{Chen:2012}
\begin{equation} q\sim\int\limits_{S^3}^{}{e^{-2\sqrt{\frac{2}{3}} \phi}\star {\cal F}},
\end{equation}
\noindent
where ${\cal F}=d{\cal A}$ is the field strength 2-form, $\star$ is the Hodge dual operator, $\phi$ is the dilaton, and $S^3$ is an arbitrary closed three-dimensional hyper-surface far enough from the black hole. The constant factor depends on
the gauge field normalization. For our solution $\phi=0$ at the spacial infinity and therefore:
\begin{equation}
q\sim\int\limits_{S^3}^{}{\star {\cal F}}.
\end{equation}
This allows one to calculate $q$ in terms of the asymptotics of the time component of the gauge potential:
\begin{equation} {\cal A}_0=-\frac{8 q}{\pi r^2} \left(1+{\cal O}\left(\frac{1}{r^2}\right)\right).
\end{equation}
For our solution:
\begin{equation}
q=\frac{\pi k^2 (1+\mu) (1+c \mu) c^2 b}{(1-c)^2 (1+c) (c+\mu)}\frac{\Delta_1 \Delta_3 \Delta_5}{\Delta_6} a_3.
\end{equation}
The dipole charge ${\cal Q}$ is another independent black ring characteristic and it equals \cite{Emparan:2004}:
\begin{equation}
{\cal Q}=\frac{1}{4 \pi} \int\limits_{S^2}^{}{\cal F},
\end{equation}
\noindent
where $S^2$ is an arbitrary 2-sphere  (or any closed surface) encircling the black ring.
The dipole charge  in our case is
\begin{equation}
{\cal Q}=k \sqrt{\frac{2\mu (b-c)}{\left(1-c^2\right)}}\Delta_1 \sqrt{\Delta_6}.
\end{equation}
The event horizon volume $S$ is defined as
\begin{equation} S=\int\limits_{S^2 \times S^1}^{}{d^3 {\cal X} \sqrt{-\gamma}},
\end{equation}
\noindent
where the integration is over the horizon surface $S^2 \times S^1$, $\gamma$ is the induced metric determinant,
which is defined as follows:
\begin{equation}
\gamma_{\alpha \beta}=\frac{\partial x^{\mu}}{\partial {\cal X}^{\alpha}} \frac{\partial x^{\nu}}{\partial {\cal X}^{\beta}} g_{\mu \nu},
\end{equation}
\noindent
where $x({\cal X})$ is a parametrization of the horizon. For our metric the event horizon volume is
\begin{equation}
S=\frac{16 \sqrt{2} \pi ^2 k^3(1+b) c \sqrt{b (c+\mu) \left(1-\mu^2\right)}}{(1-c) (1+c)^2}\frac{\Delta_1 \Delta_3 \Delta_5 \Delta_7 \Delta_8}{\sqrt{\Delta_6}}.
\end{equation}
The black ring angular velocities $\Omega_{\phi}$ and $\Omega_{\psi}$ coincide with the $\phi$- and $\psi$-components of the direction vector for rod $v=-1/c$ corresponding to the event horizon. The rod direction vector should be normalized in such manner that its time component is equal to 1\cite{Harmark:2004}, that is the rod direction has the form $V=(1,\Omega_\phi,\Omega_\psi)$.
As the result one obtains:
\begin{equation}
\Omega_{\phi}=\frac{(1-c) \sqrt{(c+\mu)}}{\sqrt{2} k (1+\mu) \sqrt{\left(1-c^2\right) b}}\frac{\sqrt{\Delta_6}}{\Delta_3 \Delta_5},
\end{equation}
\bea
\Omega_{\psi} &=& \frac{\sqrt{\mu (1-c)}}{\sqrt{2} k (1+\mu)^2(1-\mu)(1+c)^{3/2} \sqrt{(b-c)} }\frac{\sqrt{\Delta_6}}{\Delta_1 \Delta_3 \Delta_5 \Delta_7 \Delta_8}  \\ \nonumber
& \times & \left\{a_1 \left[\left(c+\mu \right)^2 \Delta_1 \left(2-\Delta_5 \right)+\left(1+c \mu \right)^2\Delta_2 \Delta_5 \right] - a_3 c \left(1+c \right) \left(1-\mu ^2\right) \Delta_1 \Delta_2\right\}.
\eea
The horizon temperature can be calculated from the condition of the absence of the conical singularity after the Wick rotation to the Euclidean signature is performed \cite{Harmark:2004}. In the present case it has the form:
\begin{equation}
T=\frac{(1-c) \sqrt{(1-\mu)}}{4 \sqrt{2} \pi k \sqrt{(1+\mu) (c+\mu) b}}\frac{\Delta_4 \sqrt{\Delta_6}}{\Delta_1 \Delta_7 \Delta_8}. \end{equation}
All these quantities were calculated for the regular case, which means that everywhere in this section $b$, $c$ and $\mu$ satisfy the condition $(\ref{m b c regularity})$. In the particular case $a_3=0$, corresponding to turning off the usual electric charge, the calculated quantities coincide with those presented in \cite{Chen:2012}. We have checked also that these quantities satisfy the First law of the black hole mechanics:
\begin{equation} \label{The First Law}
dM=\frac{1}{4} TdS+\Omega_{\phi} dJ_{\phi}+\Omega_{\psi} dJ_{\psi}+\varrho dq+\vartheta d{\cal Q} ,
\end{equation}
\noindent
where $\varrho$ and $\vartheta$ are the potentials corresponding to the electric charge and the dipole charge respectively. The First law is in our case a system of five equations. We used two of them to find $\varrho$ and $\vartheta$. Then we checked that the remaining three equations are satisfied. The result for the potentials is
\begin{equation}
\vartheta=\frac{\pi k (1-\mu)}{\sqrt{2} (c+\mu)} \sqrt{\frac{(1-c)(b-c) \mu}{\left(1+c \right)}} \frac{\Delta_4 \Delta_5 \Delta_9}{\Delta_8 \sqrt{\Delta_6}},
\end{equation}
\begin{equation}
\varrho = \frac{1}{\Delta_3 \Delta_5 \Delta_8} \frac{2 c}{c+\mu}\left[\frac{1-c}{1+c} \frac{1-\mu}{1+\mu} \left(a_3+a_1 \frac{(1-c)(c+\mu)}{(1+c)(1+\mu)c} \right) \mu \Delta_4 \Delta_6+2 c \frac{1+c \mu}{1-c^2} \Delta_3 \Delta_8 a_3\right].
\end{equation}

\section{Conclusions}
In this paper we constructed the general black ring solution in the 5D Einstein-Maxwell-dilaton theory. This model is the Kaluza-Klein compactification of the pure Einstein 6D gravity and coincides with bosonic sector of 5D $U(1)^3$ supergravity when only one of the gauge fields is non-zero. Then the regularity conditions were found and mass, angular momenta, charges and all other quantities entering the First law of black hole mechanics were calculated and the validity of the First law was checked.

The most obvious goal now could be to find the general black ring solution in $U(1)^3$ supergravity with all gauge fields nonvanishing. A possible path to this goal is to apply boosts and T-dualities to the solution presented here. Then one can try to symmetrize the result by introducing the missing parameters. Another possible continuation of this work would be to consider the compactification to 4D charged C-metric that will generalize the one found in \cite{Dowker:1993}.

 \begin{acknowledgments}
 The investigation was supported in part by the Russian
 Foundation for Basic Research through Grant No. 11-02-00792-a and by the Grant 14.740.11.0082 of Federal special-purpose
program "Scientific and scientific-pedagogical personnel of innovative Russia".
 \end{acknowledgments}

\appendix
\section{Metric and inverse metric components}
Let us list here explicitly all metric components:
\beq
g_{ww}=\frac{D(u,v)}{H(u,v)}\,,\quad g_{tw}=\frac{K(u,v)}{H(u,v)}\,,\quad g_{\phi\psi}=\frac{J(u,v)}{H(u,v)}.
\eeq
\bea
g_{\phi\phi}&=&\frac{1}{H(u,v)}\Biggl[ S\left\{\left(1-\frac{(u-y_1)^2G(v)}{(u-v)^2 G(y_1)}\right)\frac{v-y_2}{v-y_1}\right\} \\ \nonumber
&+& 2a_1 a_2 a_3 \frac{r_1 r_2 r_3}{(u-v)^2 G(y_1)}\sum_i \frac{1}{a_i r_i}S_i\left\{(u-v)^2\left(x-y_1+G(v)\frac{u-x_i}{v-x_i}\right) \right\}\Biggr]\,,
\eea
\beq
g_{t\phi}=\frac{1}{H(u,v)}\left[S\{v-y_2\}-a_1 a_2 a_3 r_1 r_2 r_3\sum_i \frac{C_i}{a_i r_i} \left(\frac{1}{y_1-x_i}- \sum_j\frac{1}{y_1-x_j}\right)\right]\,,
\eeq
\beq
g_{t\psi}=-\frac{1}{H(u,v)}\left[\sum_i \frac{a_i r_i}{(y_2-x_i)(u-v)} S^\prime_i\{(u-v)(v-x_i)\}\right]\,.
\eeq
The metric has the following symmetry. It changes sign when one exchanges simultaneously $t\leftrightarrow w$, $\phi\leftrightarrow \psi$, $u\leftrightarrow v$, $y_1\leftrightarrow y_2$. This symmetry allows one to obtain easily the components that are not written here explicitly. For example, $g_{w\psi}(u,v,y_1,y_2)=-g_{t\phi}(v,u,y_2,y_1)$. The inverse metric has the same symmetry.

The inverse metric components are
\bea
g^{tt} &=& \frac{1}{H(u,v)}\Biggl[ S\left\{\left((u-y_1)^2-(u-v)^2\frac{G(y_1)}{G(v)}\right)\frac{v-y_2}{v-y_1}\right\} \\ \nonumber
&-& 2a_1 a_2 a_3 \frac{r_1 r_2 r_3}{(u-v)^2}\sum_i \frac{1}{a_i r_i}S_i\left\{(u-v)^2\left(1-(u-v)^2\frac{(y_1-x_i)(v-x_i)}{(u-x_i)G(v)}\right) \right\}\Biggr]\,,
\eea
\bea
g^{tw}&=& \frac{1}{H(u,v)}\Biggl[-2 (u-v)^3\frac{a_1 a_2 a_3 r_1 r_2 r_3 }{G(y_1)G(y_2)}\sum_i \frac{dG}{du}\vert_{u=x_i}\\ \nonumber
&-&\sum_i \frac{a_i r_i}{(u-v)} S^\prime_i\left\{(u-v)^2\left(1-\frac{u-v}{G(u)}(u-y_1)(u-x_i)\right)\left(1+\frac{u-v}{G(v)}(v-y_2)(v-x_i)\right)\right\}\Biggr]\,.
\eea
\bea
g^{t\phi}&=&  \frac{G(y_1)}{H(u,v)}\Biggl[ S\left\{\frac{(u-v)^2}{G(v)}(v-y_2)\right\}\\ \nonumber
&-& \frac{a_1 a_2 a_3 }{(u-v)^2}\sum_i \frac{r_1 r_2 r_3}{a_i r_i} S_i\left\{\frac{(u-v)^4(v-x_i)}{G(v)(u-x_i)}\right\} \left(\frac{1}{y_1-x_i}- \sum_j\frac{1}{y_1-x_j}\right)\Biggr]\,,
\eea
\bea
g^{t\psi}&=& \frac{1}{H(u,v)}\Biggl[-2 (u-v)^3\frac{a_1 a_2 a_3 r_1 r_2 r_3 }{G(y_1)G(y_2)}\sum_i \left(y_2+\frac{1}{2}\left(x_i-\sum_j x_j\right)\right)\frac{dG}{du}\vert_{u=x_i}\\ \nonumber
&+& G(y_2)\frac{(u-v)^2}{G(u)G(v)}\sum_i \frac{a_i r_i}{(y_2-x_i)} S^\prime_i\left\{\left((y_1-u)(u-x_i)+\frac{G(u)}{u-v}\right)\left(v-x_i\right)\right\}\Biggr]\,,
\eea
\beq
g^{\psi\psi}=-(u-v)^4\frac{G(y_2)}{G(u)G(v)}\frac{F(u,v)}{H(u,v)}\,, \quad g^{\phi\psi}=(u-v)^4\frac{G(y_1)G(y_2)}{G(u)G(v)}\frac{J(u,v)}{H(u,v)}\,.
\eeq

Potential ${\cal N}^A_B$ used in the demonstration of Ricci flatness of the metric has the following components:
\bea
{\cal N}^t_t &=& \frac{1}{H(u,v)}\Biggl[ S\left\{(v-y_2)\left(u(u-y_1)-\frac{G(u)}{u-v}\right)\right\} \\ \nonumber
&-& a_1 a_2 a_3 r_1 r_2 r_3\sum_i \frac{C_i}{a_i r_i}(x_1+x_2+x_3-x_i)\Biggr]\,,
\eea
\beq
{\cal N}^t_\phi = \frac{G(u)G(v)}{G(y_1)(u-v)^4}g^{t\phi}\,, \quad {\cal N}^t_\psi = -\frac{G(u)G(v)}{G(y_1)(u-v)^4}g^{t\psi}\,,
\eeq
\beq
{\cal N}^t_w = -\frac{1}{H(u,v)}\sum_i \frac{a_i r_i}{u-v} S^\prime_i\{G(u)-(u-v)(u-y_1)(u-x_i)\}\,,
\eeq
\beq
{\cal N}^\phi_t =-G(y_1)g_{t\phi}\,, \quad {\cal N}^\phi_\phi = F(v,u)\,,\quad {\cal N}^\phi_\psi =-G(y_1)J(u,v)\,, \quad {\cal N}^\phi_w =-G(y_1)g_{w\phi}\,.
\eeq
Potential ${\cal N}^A_B$ is invariant under simultaneous replacements  $t\leftrightarrow w$, $\phi\leftrightarrow \psi$, $u\leftrightarrow v$, $y_1\leftrightarrow y_2$. This allows one to recover the other half of the components. For example, ${\cal N}^w_\psi(u,v,y_1,y_2)={\cal N}^t_\phi(v,u,y_2,y_1)$. Here we chose such normalization of $\epsilon^{ij}$, that Eq. (\ref{Ricci_flat}) takes the form:
\begin{equation}
-\frac{G(u)}{(u-v)^2} g^{A B} \partial_{u} g_{B C}=\partial_{v} {\cal N}^A_C, \quad \frac{G(v)}{(u-v)^2} g^{A B} \partial_{v} g_{B C}=-\partial_{u} {\cal N}^A_C.
\end{equation}

\end{document}